\documentstyle[12pt]{article}
\normalbaselines
\begin{document}
\bibliographystyle{unsrt}

\vbox {\vspace{6mm}}
INFN-NA-IV-94/30~~~~~~~~~~~~~~~~~~~~~~~~~~~~~~~~~~~~~~~DSF-T-94/30

\begin{center}
{\large \bf NONCLASSICAL LIGHT IN INTERFEROMETRIC MEASUREMENTS}\\[7mm]
N. A. Ansari, L. Di Fiore, R. Romano, S. Solimeno and F. Zaccaria\\
{\it Dipartimento di Scienze Fisiche,\\
Universita di Napoli "Federico II",\\
Istituto Nazionale di Fisiche Nucleare, Sezione di Napoli,\\
Mostra d'Oltremare, Pad. 20-80125 Napoli, Italy}\\
and\\
M. A. Man'ko and V. I. Man'ko\\
{\it Lebedev Physics Institute,\\
Leninsky pr.,53, 117924 Moscow, Russia}\\[5mm]
\end{center}

\vspace{2mm}

\begin{abstract}
It is shown that the even and odd coherent light and other nonclassical
states of light like superposition of coherent states with different
phases may replace the squeezed light in interferometric gravitational
wave detector to increase its sensitivity.

(Contribution to the Second Workshop on Harmonic Oscillator, Cocoyoc,
Mexico, March 1994)

\end{abstract}

\section{Introduction}
The problem of detecting gravitational wave has been a subject of
interest for many years \cite{[1]}. Specially the quantum sensitivity of
Michelson interferometric gravitational wave detection (GWD) has been
discussed by Caves \cite{[2]}. In Michelson interferometer, the light
from an input laser beam splits through a 50-50 beam splitter (BS),
bounces back and forth between two end mirrors of interferometer and
recombines again at the BS. The intensity at one or both output ports of
the interferometer provides informations about the difference between the
two displacements of the end mirrors. The quantum mechanical treatment
of the system shows that the vacuum fluctuations enter in to the
interferometer from the unused port and result in a limit on the optimum
power of the input laser, which comes out to be quite large and of no
experimental interest. Caves \cite{[2]} suggested that by squeezing the
vacuum, the optimum power of the laser can be reduced considerably.
Squeezed states \cite{[3]} of an electromagnetic field are non-classical
states in which the quantum fluctuations in one quadrature can be
reduced below the standard quantum limit at the expanse of the increased
fluctuations in the other quadrature such that the Heisenberg uncertainty
principle remains valid.

It is also interesting to try to use the other non-classical light in the
place of squeezed light and study its effect on the better sensitivity of
the interferometer in GWD. The different superpositions of coherent states
because of their non-classical nature are of our particular interest.
Yurke and Stoler \cite{[4]}, have predicted that a coherent state
propagating in a dispersive medium evolves into a superposition of two
coherent states $180^0$ out of phase. Another type of superposition of
coherent states, namely, even and odd coherent states was introduced by
Dodonov, Malkin, and Man'ko \cite{[5]}. Even coherent states are closely
related to the squeezed vacuum states because they too are the
superposition of even number of photons but with different coefficients.
The non-classical properties of Yurke-Stoler coherent states and even and
odd coherent states have been discussed in \cite{[6]}. In Refs.\cite{[7]}
-\cite{[10]}, different theoretical possibilities regarding the generations
of even and odd coherent states have been discussed. The properties of
even and odd coherent states as a representatives of a set of nonclassical
light states have been considered recently by Nieto and Truax \cite{[11]}.

In the following sections we will study the effects of the non-classical
light on the optimal power of the input laser for interferometric GWD.
The most general analysis of non-classical states in interferometry was
done by Yurke, McCall and Klauder \cite{[12]}. We will following the
approach adopted by Ansari et al.\cite{[13]}, in which the noise error
can be expressed as a product of two factors with tensorial-like
structure, each of the factors being related to the geometry of
an interferometer and light states correspondingly.

\section{Nonclassical Light}

In this section we will briefly discuss the properties of three types of
superposition of coherent states, Yurke-Stoler coherent states (YS),
even (ECS) and odd (OCS) coherent states.

\subsection{Even and Odd Coherent States}

The even and odd coherent states may be defined in the form \cite{[5]}
\begin{equation}
\mid\beta_{\pm}>=N_{\pm}(\mid\beta>\pm\mid -\beta>),
\end{equation}
where + sign is for ECS and -- sign is for OCS. $\mid\beta>$ is a
coherent state and the normalizing constants $N_{\pm}$ are
\begin{eqnarray}
N_{+}&=&\frac{e^{\mid\beta\mid^{2}/2}}{2\sqrt{cosh\mid\beta\mid^{2}}},
\nonumber\\
N_{-}&=&\frac{e^{\mid\beta\mid^{2}/2}}{2\sqrt{sinh\mid\beta\mid^{2}}}.
\end{eqnarray}
Also from Eq.(1), we can define the relations
\begin{eqnarray}
a\mid\beta_{+}>&=&\beta\sqrt{tanh\mid\beta\mid^{2}}\mid\beta_{-}>,
\nonumber\\
a\mid\beta_{-}>&=&\beta\sqrt{coth\mid\beta\mid^{2}}\mid\beta_{+}>.
\end{eqnarray}
With the help of above equations we can easily evaluate the expectation
values of first and higher order moments of annihilation and creation
operators of even and odd coherent states. For example,
\begin{equation}
<a>_{+}=<\beta_{+}\mid a\mid\beta_{+}>=\beta\sqrt{tanh\mid\beta\mid^{2}}
<\beta_{+}\mid\beta_{-}>=0,
\end{equation}
as even and odd coherent states are orthogonal states. Similarly,
\begin{eqnarray}
<a^{\dag}a>_{+}&=&\mid\beta\mid^{2}tanh\mid\beta\mid^{2},\nonumber\\
<a^{\dag}a>_{-}&=&\mid\beta\mid^{2}coth\mid\beta\mid^{2},\nonumber\\
<a^{2}>_{\pm}&=&\beta^{2},\nonumber\\
<a^{\dag 2}>_{\pm}&=&\beta^{*2}.
\end{eqnarray}

\subsection{Yurke-Stoler Coherent States}

Yurke-Stoler (YS) coherent states are defined as \cite{[4]},\cite{[6]}
\begin{equation}
\mid\beta>_{YS}=\frac{1}{\sqrt{2}}(\mid\beta>+e^{i\pi/2}\mid-\beta>).
\end{equation}
In terms of number states these states can be defined as
\begin{equation}
\mid\beta>_{YS}=\frac{e^{-\mid\beta\mid^{2}/2}}{\sqrt{2}}
\sum_{n=0}^{\infty}\frac{\beta^{n}}{\sqrt{n!}}(1+i(-1)^{n})\mid n>.
\end{equation}
The first order moments of YS coherent states are not equal to zero as
in the case of ECS or OCS
\begin{equation}
<a>_{YS}=-i\beta e^{-2\mid\beta\mid^{2}},
\end{equation}
and second order moments are
\begin{eqnarray}
<a^{\dag}a>_{YS}&=&\mid\beta\mid^{2},\nonumber\\
<a^{2}>_{YS}&=&\beta^{2}.
\end{eqnarray}
We will use different first and second order moments as given in
Eqs.(4-9) in the following section, when we will discuss the important
role played by nonclassical light for GWD.

\section{Michelson Interferometer for GWD}

Michelson interferometer is a two arms device at the end of which two
mirrors are attached to strings, thus behaving as two pendula. The
positions of the mirrors are controlled by the joint action of the
restoring force and the radiation pressure \cite{[14]}. We will suppose
that in all process the dissipative and active effects are negligible
and the conservation of energy is ensured.

There are two input field modes described by the operators
$(a_{i},a^{\dag}_{i})$ at the two ports of the interferometer. At the
end mirrors $M_{i}$, the fields are defined by $(b_{i},b^{\dag}_{i})$.
The output fields at the two ports $P_{i}$ are described by
$(c_{i},c^{\dag}_{i})$. The input fields are related with the fields at
the mirrors through the relations
\begin{eqnarray}
b&=&Va,\nonumber\\
b^{\dag}&=&a^{\dag}V^{\dag },
\end{eqnarray}
where
\begin{eqnarray}
a&=&\left(\begin{array}{c}
a_{1}\\ a_{2} \end{array} \right),\nonumber\\
b&=&\left(\begin{array}{c}
b_{1}\\ b_{2} \end{array} \right),\nonumber\\
a^{\dag}&=&\left(a_{1}^{\dag}~~a_{2}^{\dag}\right),\nonumber\\
b^{\dag}&=&\left(b_{1}^{\dag}~~b_{2}^{\dag}\right).
\end{eqnarray}
Also
\begin{equation}
V=\Phi K,
\end{equation}
with
\begin{eqnarray}
\Phi=\left(\begin{array}{clcr}
e^{i\phi_{1}}  & 0\\
0 & e^{i\phi_{2}}\end{array}\right),\nonumber\\
K=\left(\begin{array}{clcr}
\zeta_{1} & \xi_{2}\\
\xi_{1} & \zeta_{2}\end{array}\right).
\end{eqnarray}
In Eq.(13) $\zeta$ and $\xi$ are the complex transitivity and
reflectivity parameters of the BS arbitrarily oriented for the $i$-th
field mode respectively and $\phi _{i}$ is the phase distance between
BS and $M_{i}$.

The relations between the input field and the output fields at the two
interferometric ports are of the form
\begin{eqnarray}
c&=&Ua,\nonumber\\
c^{\dag}&=&a^{\dag}U^{\dag}
\end{eqnarray}
with
\begin{equation}
U=-K^{T}\Phi^{2}K=-V^{T}V,
\end{equation}
where -- sign in Eq.(15) corresponds to the phase change on reflection
at the mirrors. Thus from the above equations we can define the
relations between different fields by including all the informations
about influence of the BS and the end mirrors $M_{i}$.

\subsection{Sources of Noise}

The accuracy with which the difference in displacement $z$ can be measured
is limited by the Heisenberg uncertainty principle. Following \cite{[2]},
we have two sources of errors namely radiation pressure error and photon
counting noise. The standard quantum limit for a Michelson
interferometer can be obtained by balancing these two sources of error.
Radiation pressure error (PR) is due to the pressure exerted by the
field on the mirrors and the photon counting noise (PC) is due to
the fluctuations in the number of photons in the input field. So,
\begin{equation}
\Delta z=\sqrt{(\Delta z_{RP})^{2}+(\Delta z_{PC})^{2}},
\end{equation}
where
\begin{eqnarray}
(\Delta z_{RP})^{2}
&=&\sigma_{RP}^{2}\left(\frac{\hbar\omega\tau}{mc}\right)^{2},\nonumber\\
(\Delta Z_{PC})^{2}
&=&\sigma_{PC}^{2}\left(\frac{\partial(c^{\dag}\sigma_{3}c)}{\partial
(\phi_{2}-\phi_{1})}\right)^{-2}.
\end{eqnarray}
Also
\begin{eqnarray}
\sigma_{RP}^{2}
&=&<(b^{\dag}\sigma_{3}b)^{2}>-<b^{\dag}\sigma_{3}b>^{2},\nonumber\\
\sigma_{PC}^{2}
&=&<(c^{\dag}\sigma_{3}c)^{2}>-<c^{\dag}\sigma_{3}c>^{2}.
\end{eqnarray}
In Eq.(17), $\tau$ is the observation time and $m$ is the mass of the
end mirrors. Here we consider that BS is attached to a large mass $M$
($M>>m$), which remained fixed during the observation time. By using
Eqs.(10-15), we can write
\begin{eqnarray}
\sigma_{RP}^{2}&=&(V^{\dag}\sigma_{3}V)_{ik}(V^{\dag}\sigma_{3}V)_{mn}
T_{ikmn},\nonumber\\
\sigma_{PC}^{2}&=&(U^{\dag}\sigma_{3}U)_{ik}(U^{\dag}\sigma_{3}U)_{mn}
T_{ikmn},
\end{eqnarray}
with the summation over the repeated indices taken from 1 to 2 and
\begin{equation}
T_{ikmn}=<a_{i}^{\dag}a_{k}a_{m}^{\dag}a_{n}>-
<a_{i}^{\dag}a_{k}><a_{m}^{\dag}a_{n}>.
\end{equation}
Eq.(20) allows us to study the use of different field modes from the
input port. By using Eqs.(16-20), we can write
\begin{equation}
\Delta z=X_{ikmn}T_{ikmn}~~~~~~~~~(ikmn=1,2),
\end{equation}
where $X_{ikmn}$ contains the geometrical and physical properties of the
interferometer.

If we consider a 50-50 ideally thin BS which introduces a phase
difference of $\pi/2$ between the reflected and the transmitted waves,
then from Eq.(10) and (13), we can write
\begin{equation}
V^{\dag}\sigma_{3}V=\left(\begin{array}{clcr}
o & i\\
-i & o \end{array}\right)
\end{equation}
and
\begin{equation}
U^{\dag}\sigma_{3}U=\left(\begin{array}{clcr}
-cos\phi & -sin\phi\\
-sin\phi & cos\phi \end{array}\right),
\end{equation}
where $\phi=\phi_{2}-\phi_{1}$. Also
if the interferometer is operated in the dark fringe, then two
arms of the interferometer can be adjusted such that $\phi=(2n+1)\pi/2$.
For dark fringe operation we get
\begin{eqnarray}
X_{1212}=X_{2121}&=&-A^{2}+B^{2},\nonumber\\
X_{1221}=X_{2112}&=&A^{2}+B^{2}.
\end{eqnarray}
Also
\begin{eqnarray}
A&=&\left(\frac{\hbar \omega \tau}{mc}\right),\nonumber\\
B&=&\left(\frac{\partial I}{\partial Z}\right)^{-1}
\end{eqnarray}
and
\begin{eqnarray}
I&=&<c^{\dag}\sigma_{3}c>,\nonumber\\
Z&=&\phi\frac{c}{2\omega}.
\end{eqnarray}
The variable $Z$ corresponds to the difference between the displacement of
two end mirrors with respect to their mean position due to radiation
pressure exerted by the input laser.

(i)~~~~The corresponding field contributions can be found from Eq.(20).
If we consider that the input field at port $P_{1}$ is a coherent light
and from the second port is in even or odd coherent states, then the two
fields are anticorrelated and the states of these fields can be written
as
\begin{equation}
\mid\psi>=\mid\alpha,\beta_{\pm}>.
\end{equation}
For the case of even coherent light we can write the coefficients
$T_{ikmn}$ as
\begin{eqnarray}
T_{1111}&=&\alpha^{2}\nonumber\\
T_{1122}&=&0\nonumber\\
T_{1212}&=&\alpha^{2}\mid\beta\mid^{2}e^{2i\theta_{1}}\nonumber\\
T_{1221}&=&\alpha^{2}\mid\beta\mid^{2}tanh\mid\beta\mid^{2}+
\alpha^{2}\nonumber\\
T_{2112}&=&\alpha^{2}\mid\beta\mid^{2}tanh\mid\beta\mid^{2}+
\mid\beta\mid^{2}tanh\mid\beta\mid^{2}\nonumber\\
T_{2121}&=&\alpha^{2}\mid\beta\mid^{2}e^{-2i\theta_{1}}\nonumber\\
T_{2211}&=&0\nonumber\\
T_{2222}&
=&\mid\beta\mid^{4}-\mid\beta\mid^{4}tanh^{2}~\mid\beta\mid^{2}+
\mid\beta\mid^{2}tanh\mid\beta\mid^{2},
\end{eqnarray}
where $\theta_{1}$ is the phase of $\beta$ and we have consider
$\alpha$ to be real. Also for OCS we will get the same expressions
as in the above equation except $tanh\mid\beta\mid^{2}$ should be
replace by $coth\mid\beta\mid^{2}$.

(ii)~~~~For the case of Yurke-Stoler coherent states from the
second port and the coherent state from the first port we can
define the states as
\begin{equation}
\mid\psi>=\mid\alpha,\beta_{YS}>,
\end{equation}
and the new expressions for $T_{ikmn}$ are
\begin{eqnarray}
T_{1111}&=&\alpha^{2}\nonumber\\T_{1122}&=&0\nonumber\\
T_{1212}&=&\alpha^{2}\mid\beta\mid^{2}e^{2i\theta_{2}}
(1+e^{-4\mid\beta\mid^{2}})\nonumber\\
T_{1221}&
=&\alpha^{2}\left[\mid\beta\mid^{2}\left(1-e^{-4\mid\beta\mid^{2}}
\right)+1\right]\nonumber\\
T_{2112}&
=&\mid\beta\mid^{2}\left[\alpha^{2}\left(1-e^{-4\mid\beta\mid^{2}}
\right)+1\right]\nonumber\\
T_{2121}&=&\alpha^{2}\mid\beta\mid^{2}e^{-2i\theta_{2}}
(1+e^{-4\mid\beta\mid^{2}})\nonumber\\
T_{2211}&=&0\nonumber\\
T_{2222}&=&\mid\beta\mid^{2},
\end{eqnarray}
where $\theta_{2}$ is the phase of $\beta$ in the case of YS coherent
states. A comparison of Eqs.(29) and (31) shows the difference between
different order correlations between the two types of the input fields
from port $P_{2}$.
\subsection{Optimum Input Laser Power}

The general expression for $(\Delta z)^{2}$ by using Eqs.(21) and (25)
becomes
\begin{equation}
(\Delta z)^{2}=A^{2}(T_{1221}+T_{2112}-T_{1212}-T_{2121})+
B^{2}(T_{1221}+T_{2112}+T_{1212}+T_{2121}).\nonumber\\
\end{equation}
Minimizing the total error with respect to $\alpha^{2}$ gives optimal
value of $\alpha^{2}$ (coherent field intensity from port $P_{1}$). In
the presence of ordinary vacuum fluctuations from the second port, the
optimum intensity of the input laser becomes \cite{[2]}
\begin{equation}
(\alpha^{2}_{opt})^{o}=\frac{mc^{2}}{2\hbar\omega^{2}\tau}.
\end{equation}
Caves \cite{[2]} showed that the optimal laser intensity can be reduced
considerably if we squeezed the vacuum from the second port. We will
analyze the situation when the squeezed vacuum is replaced by the
nonclassical light as discussed before.

In the first case, we will study the effect of even and odd coherent
states on the optimum value of $\alpha^{2}$. Under the condition of
$\alpha^{2}>>\mid\beta\mid^{2}tanh\mid\beta\mid^{2}$, we get
\begin{equation}
(\alpha^{2}_{opt})^{ev}
=\sqrt{\frac{2\mid\beta\mid^{2}tanh\mid\beta\mid^{2}+
2\mid\beta\mid^{2}cos2\theta_{1}+1}
{2\mid\beta\mid^{2}tanh\mid\beta\mid^{2}-
2\mid\beta\mid^{2}cos2\theta_{1}+1}}(\alpha^{2}_{opt})^{o},
\end{equation}
and for OCS
\begin{equation}
(\alpha^{2}_{opt})^{od}
=\sqrt{\frac{2\mid\beta\mid^{2}coth\mid\beta\mid^{2}+
2\mid\beta\mid^{2}cos2\theta_{1}+1}
{2\mid\beta\mid^{2}coth\mid\beta\mid^{2}-
2\mid\beta\mid^{2}cos2\theta_{1}+1}}(\alpha^{2}_{opt})^{o}.
\end{equation}
Thus for $\theta_{1}=\pi/2$ and under the limit
$1<<\mid\beta\mid^{2}<<\alpha^{2}$, we get
\begin{equation}
(\alpha^{2}_{opt})^{ev}=\frac{(\alpha^{2}_{opt})^{o}}{2\mid\beta\mid}.
\end{equation}
Eq. (35) allows us an alternative way to reduce the optimum input
laser power or to increase the sensitivity of interferometer by using
even or odd coherent states from the second port of the interferoeter. As
$\mid\beta\mid>>1$, from Eq.(35), we predict that the optimum value of
the input laser intensity can be reduced considerably if we apply even or
odd coherent state from the second port.

When we apply Yurke-Stoler coherent states and for the choices of
$\alpha^{2}>>\mid\beta\mid^{2}$ and $\theta_{2}=\pi/2$, we get the
relation
\begin{equation}
(\alpha^{2}_{opt})^{YS}
=\sqrt{\frac{-2\mid\beta\mid^{2}e^{-4\mid\beta\mid^{2}}
+1}{4\mid\beta\mid^{2}+1}}(\alpha^{2}_{opt})^{o}.
\end{equation}
Also in the limit of $1<<\mid\beta\mid^{2}<<\alpha^{2}$, we will get
the same expression as we get in the case of ECS or OCS, i.e.,
\begin{equation}
(\alpha^{2}_{opt})^{YS}=\frac{(\alpha^{2}_{opt})^{o}}{2\mid\beta\mid}.
\end{equation}
Eqs.(35) and (37) show that we get the same expressions for the optimum
power of input laser for large $\mid\beta\mid$. Thus we predict an
important property of the superposition of coherent states that differet
superpositions of coherent states may play an important role in reducing
the optimum power of input laser. In other words by applying these
coherent states, better quantum sensitivity of interferometer can be
achieved as compare to no field contribution from the second port.

\end{document}